\documentstyle[prl,aps,amssymb,twocolumn]{revtex}

\input BoxedEPS
\SetRokickiEPSFSpecial  %% for dvips by Tom Rokicki, for VMS
\HideDisplacementBoxes

\begin{document}

\draft

\title {Kolmogorov-Sinai Entropy-Rate vs.\ Physical Entropy}

\author{ Vito Latora\cite{AVL} and Michel Baranger\cite{AMB}}

\address{Center for Theoretical
Physics, Laboratory for Nuclear Sciences
and Department of Physics,
\\ Massachusetts Institute
of Technology, Cambridge, Massachusetts 02139, USA\\[1ex]
\rm MIT CTP\#2751  \qquad chao-dyn/9806006}

%\date{\today}
\maketitle

\begin{abstract}
This letter elucidates the connection between the KS entropy-rate $\kappa$ and
the time evolution of the physical or statistical entropy $S$.  For a large
family of chaotic conservative dynamical systems including the simplest ones,
the evolution of $S(t)$ for far-from-equilibrium processes includes a stage
during which $S$ is a simple linear function of time whose slope is $\kappa$.
The letter presents numerical confirmation of this connection for a number
of chaotic symplectic maps, ranging from the simplest 2-dimensional ones to a
4-dimensional and strongly nonlinear map.
\end{abstract}

\pacs{05.45.+b, 05.70.Ln}

This paper tries to clarify the connection between the Kolmogorov-Sinai
entropy and the physical entropy for a chaotic conservative dynamical system.
This connection is obviously very important if one is to understand the impact
on thermodynamics and statistical mechanics of the large amount of work done by
mathematicians on the behavior of chaotic systems.  To start with the KS
entropy, it is not really an entropy but an entropy per unit time, or an
``entropy-rate''.  It is a single number $\kappa$, which is a property solely
of the chaotic dynamical system considered. As for the physical entropy $S(t)$,
the entropy of the second law of thermodynamics, it is a function of time, and
this function depends not only on the particular dynamical system, but also on
the choice of an initial probability distribution for the state of that system.
Though it is clear that the original definition of $\kappa$~\cite{kolmo} was
meant to provide a connection with $S(t)$, the precise connection does not seem
to be well known nowadays, and the few statements found in the textbooks are
often vague ~\cite{zaslav}.

The simplest connection one might guess would be this: the KS entropy-rate
would be the maximum possible absolute value of the rate of variation of the
physical entropy, i.e. $|dS/dt| \le \kappa$. But this is wrong, because
a counterexample can easily be found \cite{baranger2,neel}. The actual
connection is less direct and, in many cases, it requires that $S(t)$ be
averaged over many histories (or trajectories), so as to give equal weights to
initial distributions from all regions of phase space.  Then, assuming these
initial distributions to be very far from equilibrium, the variation
with time of the physical entropy goes through three successive,
roughly separated stages. In the first stage, $S(t)$ is heavily dependent
on the details of the dynamical system and of the initial distribution;
no generic statement can be made;  $dS/dt$ can be positive
or negative, large or small, and in particular it can be larger than
$\kappa$. In the second stage, $S(t)$ is a linear increasing function whose
slope is $\kappa$. In the third stage, $S(t)$ tends asymptotically toward the
constant value which characterizes equilibrium, for which the distribution
is uniform in the available part of phase space.  It may happen, however,
that the simple and generic stage 2 is absent, with stages 1 and 3 merging
into each other. This is true when the initial distribution is not
sufficientely different from the equilibrium distribution.

We make no claim of having a rigorous mathematical proof of these
statements. We do have an incomplete, but quite suggestive, analytical
discussion \cite{baranger2}, which cannot fit in the space available here.
The latter part of this letter will present a few very convincing numerical
simulations for symplectic maps of 2 and 4 dimensions.  Ref. \cite{baranger2}
contains more map results, as well as a 3-dimensional flow.  Some of our
ideas are already present in ref. \cite{gu},
including the three-stage idea; but the crucial connection with
the KS entropy-rate, valid for any number of
dimensions and for nonlinear systems, is not there.  Note that the definition
of a single global $\kappa$ is not always a useful one, for instance for
several weakly coupled subsystems;  in such cases the connection clearly needs
generalization.

The definition of the KS entropy-rate can be found in many textbooks
\cite{billing}.
To calculate it here, we use the fact that it is equal to the sum of the
positive Lyapunov exponents \cite{pesin}. Our definition
for the out-of-equilibrium physical entropy is $S =$ constant $- I$,
where $I$ is the Shannon information \cite{balian}.
These $S$ and $I$ are coarse grained \cite{coarse}.
Coarse-graining consists in performing a slight smearing, or smoothing,
of the probability distribution in phase space before calculating
$S$ or $I$. The fine-grained quantities do not vary with time at all,
because Liouville's theorem says that the volume of phase space is
conserved. The shape of that volume, however, becomes increasingly
complicated and fractalized, due to the chaotic dynamics.
Hence, under smoothing, the volume occupied keeps increasing.
There are many ways to perform a coarse-graining.
For this paper, we assume that phase space is divided into a
large number of cells $c_{\alpha}$ with volumes
$v_{\alpha}$, such that ${\sum}_{\alpha} v_{\alpha}=V$,
the total volume of available
phase space. Then we define $I$ by
\begin{equation}
\label{scoarse}
I(t)  =        \sum_{\alpha} ~  p_{\alpha}(t) ~
             \log \left[ \frac{V}{v_{\alpha}} p_{\alpha}(t)\right]~~,
\end{equation}
where $p_{\alpha}(t)$ is the probability that the state
of the system in phase space at time $t$ falls inside
cell $c_{\alpha}$. In the following it will be more convenient to work with
$I(t)$ rather than $S(t)$. This is because, when $I$ is used, there is a
convenient reference point, the uniform distribution, whose $I$
always vanishes, irrespective of the coarse-graining. On the other hand
the constant quantity $I+S$, and therefore $S$, do depend on
the coarse-graining.

This type of coarse-graining allows an alternative version of the significance
of $\kappa$ for the evolution of a physical system.  Let us assume the initial
distribution to be very strongly localized in phase space, i.e. most cells
contain zero probability initially.  Then, during the generic ``second stage''
mentioned earlier, the total number of occupied cells, i.e. cells with
non-vanishing $p_{\alpha}$, varies proportionally to $e^{\kappa t}$. Our
simulations verify this fact well (see fig.~3 later).

We return to the need for averaging, in our simulations, many histories
starting from different parts of phase space.  This has to be done whenever the
local $\kappa$ (the sum of the positive local Lyapunov exponents) varies
appreciably from place to place, which is the normal case for nonlinear
systems. For linear maps (like the generalized cat map below) it is not
necessary. For other systems, it would never be necessary if we could use a
fine enough coarse-graining, to give the probability time to spread
throughout phase space before any appreciable increase in entropy.
Unfortunately such fine grain would require computers far more powerful than
exist now.  In the real thermodynamical world with many many dimensions, what
kind of coarse-graining should preferably be used is, we believe, a wide open
question.

For what may be the simplest of all conservative chaotic systems,
the baker's map, the correctness of our three-stage description for
the behavior of $I(t)$ can be shown analytically \cite{baranger2}. Our
first simulations are done with the ``generalized cat map'' inside a
unit square:
\begin{eqnarray}
\label{cat1}
P&=& p    + kq\pmod1,\nonumber\\
Q&=& p  +(1+k)q\pmod 1
\end{eqnarray}
where $k$ is a positive control parameter.
Fig.~1 shows $I(t)$ for four values of $k$ (see caption).
The coarse-graining grid is obtained by
dividing each axis into 400 equal segments. The initial distribution
consists of $10^6$ points placed at random inside a square whose size is that
of a coarse-graining cell, and the center of that square is
picked at random anywhere on the map. Each of the four curves is an
average over 100 runs, i.e. 100 histories with different initial
distributions chosen at random, as mentioned. Each curve shows clearly
the stage-2 linear behavior, the negative of the slope being
accurately given by the (analytically calculable) Lyapunov exponent:
\begin{equation}
\label{cat2}
\lambda= \log \frac{1}{2}({2 +k +\sqrt{k^2+4k}}) = \kappa~~.
\end{equation}

Fig.~2 shows how $I(t)$ depends on the initial distribution  
and on the coarse-graining. 
Now $k=1$ only.  Other conditions are the same as in fig.~1, except that we
calculate a single history instead of averaging 100.  This makes no big
difference in this case, because the local Lyapunov exponent is the same
everywhere. 
In the six top curves six different sizes are 
compared for the initial square; the grid is as in fig.~1.  
The first size is that of 1 coarse-graining cell,  
then 4 cells, 16, 64, 256, and 1024 cells (from top to bottom). 
All six curves have the same stage-2 slope given by 
$-\kappa$. Their vertical displacement is $\log 2$
for each factor 2 in the linear dimension of the initial distribution.
The three bottom curves show how $I(t)$ depends on the coarse-graining. 
The size of the initial square is always that of 1024 original 
coarse-graining cells. 
For the upper curve (sixth from the top) the cells are as in fig.~1, 
for the middle curve they are squares
four times larger in area, and for the bottom curve they 
are four times larger again. 
Once again, all three curves have 
the same stage-2 slope of $-\kappa$. They are displaced vertically from each
other by the log of the factor in coarse-graining linear
dimension, i.e. $\log 2$.

In fig.~3 we plot on a log scale the number of occupied cells vs.\ time for the
finest coarse-graining and 3 progressively larger initial distributions, whose
centers are picked at random as always. All 3 stage-2 straight lines are
indeed fitted by $e^{\kappa t}$. Their vertical displacement is by a
factor 4, which is also the factor in the linear size of the initial
distributions.

The second system studied is the standard map \cite {stand}, again a
two-dimensional conservative map in the unit square, but this time
nonlinear:
\begin{eqnarray}
P&=&p + \frac{k}{2\pi}\sin(2\pi q)\pmod1~,\nonumber\\
Q&=&q + P\pmod1~.
\label{sta}
\end{eqnarray}
The map is only partially chaotic, but the percentage of chaos increases
with the control parameter $k$, and we use large values of $k$, namely
20, 10, and 5. For $k=5$ there are two sizeable regular islands,
associated with a period 2 stable trajectory. We calculated the Lyapunov
exponent numerically, leaving out the regular islands for $k=5$. This
yielded $\lambda= \kappa=$ 2.30,~1.62,~0.98, respectively for the three $k$'s.
Fig.~4 shows the three curves $I(t)$, with the top curve corresponding
to the smallest $k$. The coarse-graining grid, the choice of initial
distribution, and the averaging are the same as for fig.~1, but it was
necessary to include 1000 histories in the averaging for $k=5$.
Each curve has a stage-2 linear portion whose slope is correctly
given by $-\kappa$. Fig.~5 presents 3 single histories (x's)
~for $k=5$, as well as the average curve from fig.~4 (circles).
For such a very nonlinear system, with too coarse a grain, the single curves
vary wildly and the averaging is essential.

Our next example is a four-dimensional system,
a   generalized cat map. It is a
linear symplectic map \cite{arnold}, reduced to a unit-size
hypercube by introducing ``mod 1'' in each of the four transformation
equations. The two positive Lyapunov exponents $\lambda_1$ and $\lambda_2$
can be calculated analytically, and the KS entropy-rate is
$\kappa=\lambda_1+\lambda_2$. For this system we made up the
coarse-graining grid by dividing each of the four axes into 20
equal segments. The initial distribution consists of $10^6$ points
placed at random inside one hypercube of size
${(400)}^{-1} \times {(400)}^{-1}\times {(400)}^{-1} \times {(400)}^{-1}$,
and the center of the hypercube is picked at random anywhere on the map.
Fig.~6 shows $I(t)$ for 2 single histories. They differ greatly in their
stage~1, but both have nearly linear stage~2's with the correct slope given
by $-\kappa$.

Finally we have a 4-dimensional nonlinear map, made up of two coupled standard
maps~:
\begin{eqnarray}
 %\label{sta1}
P_1&=&p_1  + \frac{k_1}{2\pi}\sin(2\pi q_1)\pmod 1\nonumber\\
Q_1&=&q_1  + P_1 + mP_2 ~~~\pmod 1  \nonumber\\
P_2&=&p_2  + \frac{k_2}{2\pi}\sin(2\pi q_2)\pmod 1\nonumber\\
Q_2&=&q_2  + P_2 + mP_1\pmod 1~.
\label{sta2}
\end{eqnarray}
We worked with 3 sets of control parameters $(k_1,k_2)$, namely (10,5), (5,3),
and (3,1).  The coupling parameter $m$ was 0.5 in all cases.  We calculated
the two Lyapunov exponents numerically \cite{vito}, then added them up to get
$\kappa$.  The numerical values are in the figure caption.  The initial volume
was the size of one cell, and we averaged over
100 histories.  Fig.~7 shows $I(t)$.  It has a fairly well defined second stage
with a slope close to $-\kappa$ in all cases.

The work we have reported makes very explicit the connection between the KS
entropy-rate, when it is meaningful, and the time dependence of the physical
entropy or information.  Yet it is far from a complete answer.  It assumes
that, at the beginning of the system's evolution, the probability distribution
spreads very fast to all corners of phase space before undergoing much
fractalization, so that its subsequent variation is well described by a global
$\kappa$.  But there are other possible scenarios.  For instance there may be
gradual overall evolution in space as well as in time, a possibility which is
not included by our assumption that all parts of phase space must start on an
equal footing.  In conclusion, although this work constitutes only a quick
foray into the subject, we hope that our assertions can function as guiding
principles for research attempting to bring together the mathematics of chaos
and the physics of far-from-equilibrium thermodynamics.

\section*{Acknowlegments}
We thank A. D'Andrea, A. Rapisarda, and M. Saraceno
for fruitful discussions.  Also S. Ganguli
and B. M\"uller who furnished some pointers to the literature.  Financial
support was provided by INFN (for V. L.), and at MIT by the U.S. Department of
Energy (D.O.E.) under contract \#DE-FC02-94ER40818.

%%%%%%%%%%%%%%%%%%%%%%%%%%%%%%%%%%%%%%%%%%%%%%%
%%%%%%%%   FIGURES %%%%%%%%%%%%%%%%%%
%%%%%%%%%%%%%%%%%%%%%%%%%%%%%%%%%%%%%%%%%%%%%%%
\begin{figure}
$$\BoxedEPSF{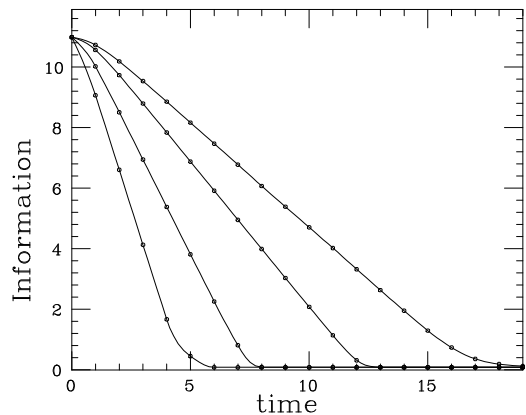 scaled 366}$$
\caption{Generalized Cat Map, $k=10$,3,1,0.5 (from left to right),
 $\kappa=2.48,1.57,0.96,0.69$ respectively,
 $N=10^6$,
 grid $=400\times400$,
 $V_i=V_{\rm cell}$,
 average of 100 histories.}
\label{fig:1}
\end{figure}

\newpage

\begin{figure}
\vspace*{-3pc}
$$\BoxedEPSF{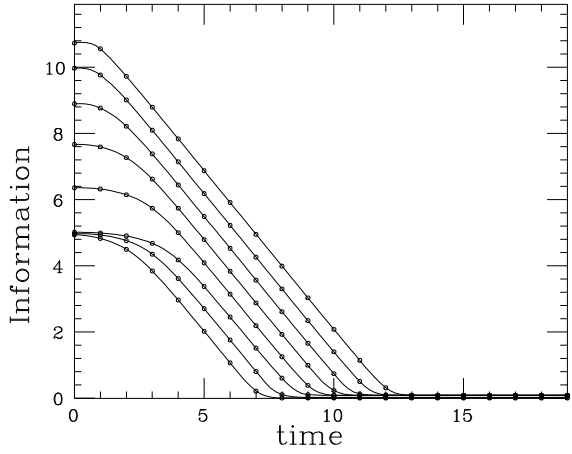 scaled 400}$$
\caption{Generalized Cat Map, $k=1,
 \kappa=0.96$,
 $N=10^6$;
 from top to bottom: $\hbox{grid} =400\times400$,
 $V_i/V_{\rm cell}=1$, 4, 16, 64, 256, 1024 (six curves), and 
 grid $=200\times200,100\times100$, 
 $V_i/V_{\rm cell}=1024$ (two curves),
 1 history.}
\label{fig:2}
\end{figure}

\begin{figure}\vspace*{-2pc}
$$\BoxedEPSF{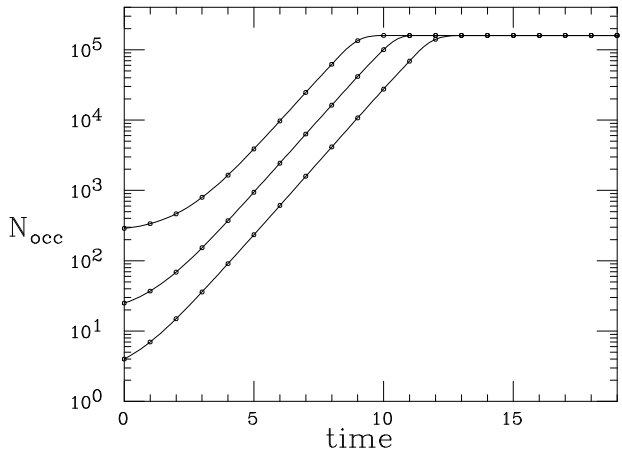 scaled 400}$$
\caption{Generalized Cat Map, $k=1, \kappa=0.96$,
 $N=10^6$,
 grid $=400\times400$,
 $V_i/V_{\rm cell}=1,16,256$ (bottom to top),
 1 history.}
\label{fig:3}
\end{figure}

\begin{figure}\vspace*{-2pc}
$$\BoxedEPSF{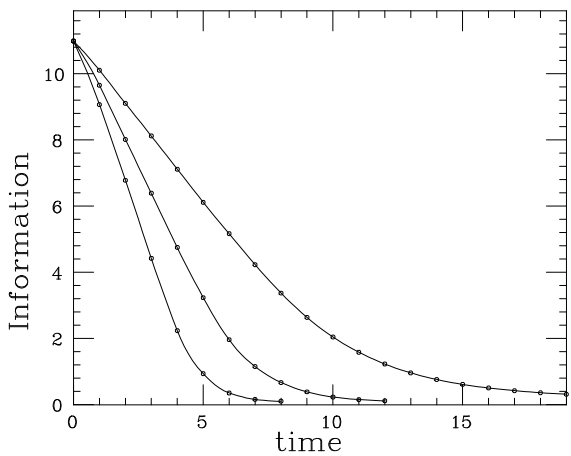 scaled 400}$$
\caption{Standard Map, $k=20,10,5$ (from left to right),
 $\kappa=2.30, 1.62, 0.98$ respectively,
 $N=10^6$,
 grid $=400\times400$,
 $V_i=V_{\rm cell}$,
 average of 100 hist.($k=20,10$), 1000 hist.\allowbreak($k=5$).}
\label{fig:4}
\end{figure}

\begin{figure}
\vspace*{-3pc}
$$\BoxedEPSF{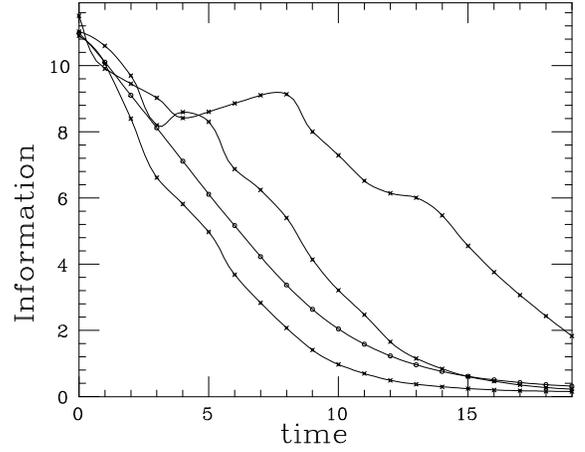 scaled 400}$$
\caption{Standard Map, $k=5$, $\kappa=0.98$,
 3 single histories compared with the average
from fig.~4.}
\label{fig:5}
\end{figure}

\begin{figure}
\vspace*{-2pc}
$$\BoxedEPSF{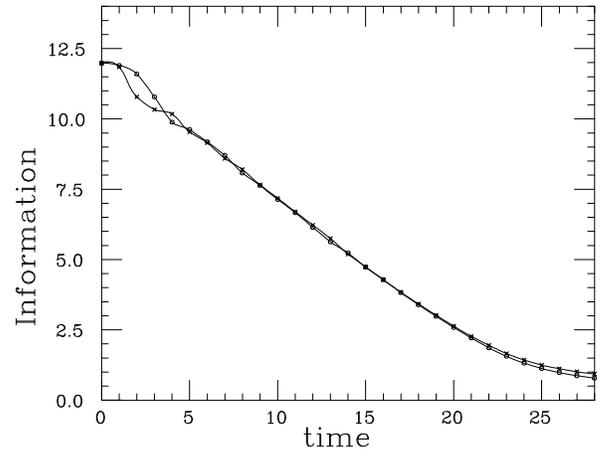 scaled 400}$$
\caption{4-Dimensional Generalized Cat Map,
 $\lambda_1=0.223$,
 $\lambda_2=0.247$,
 $\kappa=0.470$,
 $N=10^6$,
 grid $=20^4$,
 $V_i=(20)^{-4}V_{\rm cell}$,
 2 single histories.}
\label{fig:6}
\end{figure}

\begin{figure}
\vspace*{-2pc}
$$\BoxedEPSF{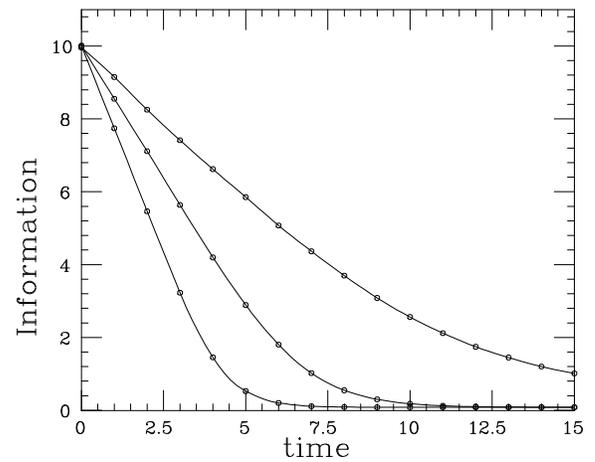 scaled 400}$$
\caption{4-Dimensional Nonlinear Map.
 Bottom: $k_1=10$, $k_2=5$, $m=0.5$,
 $\lambda_1=1.65$, $\lambda_2=0.75$, $\kappa=2.40$.
 Middle: $k_1=5$, $k_2=3$, $m=0.5$,
 $\lambda_1=1.03, \lambda_2=0.45, \kappa=1.48$ .
 Top: $k_1=3, k_2=1, m=0.5$,
 $\lambda_1=0.62, \lambda_2=0.16, \kappa=0.78$ .
 $N=10^6$,
 Grid $=20^4$,
 $V_i=V_{\rm cell}$,
 average of 100 histories.}
\label{fig:7}
\end{figure}

\end{document}